\documentclass[french,english,aps,manuscript]{revtex4}
\usepackage[LGR,T1]{fontenc}
\usepackage{textcomp}
\usepackage[latin9]{inputenc}
\usepackage{babel}
\usepackage{float}
\usepackage{mathrsfs}
\usepackage{amsmath}
\usepackage{amssymb}
\usepackage{graphicx}
\usepackage[a4paper]{geometry}
\usepackage[]
 {hyperref}

\makeatletter

\providecommand{\tabularnewline}{\\}
\newcommand{\lyxdot}{.}

\@ifundefined{textcolor}{}
{%
 \definecolor{BLACK}{gray}{0}
 \definecolor{WHITE}{gray}{1}
 \definecolor{RED}{rgb}{1,0,0}
 \definecolor{GREEN}{rgb}{0,1,0}
 \definecolor{BLUE}{rgb}{0,0,1}
 \definecolor{CYAN}{cmyk}{1,0,0,0}
 \definecolor{MAGENTA}{cmyk}{0,1,0,0}
 \definecolor{YELLOW}{cmyk}{0,0,1,0}
}

\usepackage{tikz,xcolor,hyperref}

\definecolor{lime}{HTML}{A6CE39}

\AtBeginDocument{

}

\makeatother

\begin{document}
\title{\texttt{THERMAL PROPERTIES OF GAUGE-INVARIANT GRAPHENE IN NONCOMMUTATIVE
PHASE-SPACE }}
\author{{\normalsize Ilyas Haouam} \href{https://orcid.org/0000-0001-6127-0408}{\includegraphics{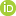}}}
\email{ilyashaouam@live.fr ; ilyas.haouam@usthb.edu.dz}

\address{Theoretical Physics and Didactic Laboratory, Faculty of Physics, University
of Sciences and Technology Houari Boumediene (USTHB), B.P. 32, El
Alia, 16111 Algiers, Algeria. }
\begin{abstract}
We study graphene in an external magnetic field within a noncommutative
(NC) framework. A gauge-invariant NC Hamiltonian is derived, and the
system is analyzed using the ladder-operator formalism, yielding deformed
Landau levels and eigenstates. A mapping to the anti-Jaynes--Cummings
model is established, providing a bridge to quantum-optical interpretation.
The thermal properties of gauge-invariant NC graphene are then investigated
via the partition function, constructed using Euler and zeta functions.
Analytical expressions for the partition function, free energy, internal
energy, entropy, and specific heat are obtained and numerically evaluated.
The results show that the NC phase-space deformation modifies the
effective Landau-level spacing and consequently alters the thermal
behavior of the graphene system. In particular, the deformation suppresses
the thermal accessibility of excited states and produces measurable
deviations from the commutative case while preserving the correct
low- and high-temperature asymptotic limits.

{\normalsize$\phantom{}$ }{\normalsize\par}

\end{abstract}
\maketitle

\section{{\normalsize Introduction}}

Graphene is a two-dimensional (2D) carbon material with a hexagonal
honeycomb lattice whose low-energy charge carriers behave as massless
Dirac quasiparticles \citep{key-1,key-2}. Since its experimental
isolation in 2004 \citep{key-3}, graphene has attracted considerable
attention due to its exceptional electronic, transport, optical, mechanical,
and thermal properties, including ultrahigh carrier mobility and long
mean free paths at room temperature \citep{key-4,key-5,key-6,key-7,key-8,key-9}.
These features make graphene a unique platform for both fundamental
research and technological applications \citep{key-10,key-11,key-12,key-13,key-14,key-15,key-16,key-17}.
Moreover, the relativistic-like electronic structure of graphene enables
tabletop studies of phenomena usually associated with high-energy
physics \citep{key-18,key-19,key-20,key-21,key-22,key-23}, while
also serving as the building block of other carbon allotropes \citep{key-24}.
In the presence of an external magnetic field (MF), the electronic
motion in graphene becomes quantized into relativistic Landau levels,
leading to several unusual quantum and thermal phenomena. In this
context, thermodynamic quantities such as the free energy, internal
energy, entropy, and specific heat provide important information about
the statistical behavior and thermal response of Dirac quasiparticles.
Since these quantities depend directly on the Landau-level spectrum,
they constitute sensitive probes of modifications induced by external
fields, quantum effects, and deformed quantum frameworks.

In parallel, noncommutative (NC) geometry, also referred to as noncommutativity
(NCy), has emerged as an important framework for modeling physics
at short length scales, originating from early ideas of Heisenberg
and Snyder \citep{key-25,key-26} and later developed systematically
by Connes and collaborators \citep{key-27,key-28,key-29}. NC formulations
have found wide applications in quantum mechanics \citep{key-30},
quantum field theory \citep{key-31}, string theory \citep{key-32},
and related areas \citep{key-33,key-34,key-35,key-36,key-37}, and
have been employed to describe diverse physical effects such as the
quantum Hall effect \citep{key-38}, the anomalous Zeeman effect \citep{key-39},
Aharonov--Bohm effect \citep{key-40}, Berry phase \citep{key-41},
and atomic spectra \citep{key-42}. They have  also been successfully
combined with standard analytical methods, including the Nikiforov--Uvarov
method \citep{key-43}, the Foldy--Wouthuysen transformation \citep{key-44},
\ensuremath{\mathscr{P}}\ensuremath{\mathscr{T}}-symmetry \citep{key-45},
spherical harmonics \citep{key-46}, the divergence theorem \citep{key-47},
Ehrenfest\textquoteright s theorem \citep{key-48}, and the WKB approximation
\citep{key-49} to solve modified eigenvalue problems in deformed
systems. Various NC structures and mappings between commutative and
NC spaces have been developed, including the $\star$product (Moyal--Weyl
product) \citep{key-50}, Seiberg--Witten (SW) map \citep{key-32},
Weyl--Wigner map \citep{key-51}, and Bopp-shift \citep{key-52}.

In this work, we investigate graphene in NC phase-space (NCPS). Our
main objective is to derive a gauge-invariant formulation of graphene
in NCPS and to analyze its thermal properties. The issue of gauge-invariance
arises for massless charged fermions coupled to an electromagnetic
field in NC space when the Dirac equation is treated within NC quantum
mechanics using only the Bopp-shift or the $\star$product \citep{key-53,key-54}.
This issue can be resolved within an NC field-theoretic framework
that combines the SW map with the $\star$product, yielding a manifestly
gauge-invariant commutative equivalent \citep{key-55,key-56}. The
study of graphene in NCPS was first considered in Ref. \citep{key-57};
however, spatial NCy could not be incorporated due to the resulting
non-gauge-invariant Dirac Hamiltonian. This leads to gauge-dependent
expressions for the particle velocity and the classical Lorentz force.
In particular, the presence of the $\Theta$-correction term prevents
the Hamiltonian from being manifestly gauge-invariant, making it unsuitable
for studying the relativistic Landau problem in graphene. Consequently,
this study restricted the analysis to momentum NCy only. Related studies
on the thermal and magnetic properties of graphene in NCPS were likewise
restricted to momentum NCy, while spatial NCy was omitted from the
analysis \citep{key-58,key-59}. These results demonstrate that,
for charged particles in NC space coupled to NC gauge fields, the
standard NC quantum-mechanical approach provides an incomplete description
with respect to gauge-invariance. Although a gauge-invariant formulation
of graphene in NC space was developed in Ref. \citep{key-55}, it
does not address the full NCPS. In contrast, the present work provides
a consistent and gauge-invariant treatment of graphene in NCPS, together
with a systematic study of its thermal behavior. This study builds
upon our recent investigations of graphene in an energy-dependent
NC and fractional frameworks \citep{key-60}. Several other works
have explored graphene within different deformed frameworks, including
Ref. \citep{key-61}, where graphene was studied in a dynamical NC
space. Another investigation of graphene in a generalized NCPS algebra,
employing a different method to introduce NC coordinates into quantum
mechanics and incorporating non-Abelian gauge fields, was presented
in Ref. \citep{key-62}. Further studies of graphene in NC space can
be found in in Refs. \citep{key-63,key-64,key-65,key-66}. In the
presence of a strong perpendicular MF, the dynamics of charged particles
confined to two dimensions can lead to an effective NC structure of
the guiding-center coordinates. This motivates the use of NCPS as
an effective framework for describing quantum corrections to graphene\textquoteright s
Landau level structure. Moreover, NC geometry has been extensively
used as a phenomenological tool to encode minimal length effects and
quantum gravitational corrections in low-energy systems. In this work,
we exploit this framework to investigate possible corrections to the
thermodynamic properties of graphene. Such deformations may also be
interpreted as effective signatures of underlying lattice-scale physics
or high-field quantum geometry effects not captured by the standard
continuum model. 

More generally, graphene physics has become a vibrant area of research,
including studies of graphene thermodynamics and related quantum phenomena;
for recent developments and references on these topics, see Refs.
\citep{key-67,key-68,key-69}. Recent progress has also extended
to deformed and generalized graphene frameworks \citep{key-70,key-71,key-72,key-73},
as well as to applications involving artificial intelligence and machine
learning techniques \citep{key-74,key-75,key-76,key-77}.

This study is organized as follows. Sec.\ref{II} explores gauge-invariant
NC graphene. Subsec.\ref{II-A} reviews the NC formalism, whereas
Subsec.\ref{II-B} constructs the gauge-invariant graphene in NCPS
and derives its eigensystem. Sec.\ref{III} investigates the thermal
properties of the deformed graphene system. Subsec.\ref{III-A} presents
the methodology and numerical results, and Subsec.\ref{III-B} provides
graphical results and discussion. Finally, Sec.\ref{IV} concludes
the paper. 

\section{{\normalsize\label{II}GAUGE-INVARIANT GRAPHENE IN NCPS}}

\subsection{{\normalsize\label{II-A}}Brief Overview of the NCPS}

At the string (ultrashort) scale, both position and momentum coordinates
become NC, rendering operator ordering nontrivial. In a 2D NCPS, the
coordinate and momentum operators, denoted by $x_{i}^{nc}$ and $p_{i}^{nc}$,
obey the following commutation relations \citep{key-30}: 
\begin{equation}
\begin{array}{cccc}
\left[x_{i}^{nc},x_{j}^{nc}\right]=\textrm{i}\Theta_{ij},\: & \left[p_{i}^{nc},p_{j}^{nc}\right]=\textrm{i}\eta_{ij},\: & \left[x_{i}^{nc},p_{j}^{nc}\right]=\textrm{i}\hbar^{eff}\delta_{ij} & ,\:i,j=1,2,\end{array}\label{eq:1}
\end{equation}
where $\Theta_{ij}=\textrm{i}\epsilon_{ij}\Theta$ and $\eta_{ij}=\textrm{i}\epsilon_{ij}\eta$
are real antisymmetric constant matrices, with $\epsilon_{ij}$ is
the Levi-Civita symbol. $\hbar^{eff}=\left(\hbar+\frac{\Theta\eta}{4\hbar}\right)$
is the effective Planck constant, and $\delta_{ij}$ is the Kronecker
delta. $\Theta$ and $\eta$ denote the spatial and momentum NC parameters,
which are fixed real-valued with the dimension of $\textrm{L}{}^{2}$
and $\textrm{M}{}^{2}\textrm{L}{}^{2}\textrm{T}{}^{-2}$, respectively,
and are assumed to be extremely small. It should be emphasized that
in our configuration, the NC parameters are typically fixed real constants.
In other types of NC frameworks, these parameters may depend on spatial
coordinates \citep{key-78}, hypermomentum \citep{key-79}, energy
\citep{key-60}, or even on time \citep{key-80}.  The NC operators
$x_{i}^{nc}$, $p_{j}^{nc}$ can be expressed in terms of the ordinary
canonical variables $x_{i}$, $p_{j}$ of standard quantum mechanics
using the Bopp-shift linear transformation as follows \citep{key-30,key-52}:
\begin{equation}
x_{i}^{nc}=x_{i}-\frac{\Theta_{ij}}{2\hbar}p_{j},\:p_{i}^{nc}=p_{i}+\frac{\eta_{ij}}{2\hbar}x_{j}.\label{eq:2}
\end{equation}

Within the NC quantum-mechanical framework, the effects of NCy can
equivalently be incorporated by replacing the ordinary pointwise product
of functions with the $\star$product. Under this deformation, a conventional
quantum system is promoted to its NC counterpart. Accordingly, if
$\mathscr{H}\left(x,p\right)$ denotes the Hamiltonian operator in
ordinary quantum mechanics, the Schrödinger equation on NC quantum
mechanics typically takes the form
\begin{equation}
\mathscr{H}\left(x,p\right)\star\psi=E^{nc}\psi.\label{eq:3}
\end{equation}

The definition of the $\star$product between two arbitrary functions
$\mathscr{F}\left(x,p\right)$ and $\mathscr{G}\left(x,p\right)$
in phase-space is given by \citep{key-30}:
\begin{equation}
\begin{array}{l}
\left(\mathscr{F}\star\mathscr{G}\right)\left(x,p\right)=\exp\left\{ \frac{\textrm{i}\Theta_{ab}}{2}\partial_{x_{a}}^{x}\partial_{x_{b}}^{x}+\frac{\textrm{i}\eta_{ab}}{2}\partial_{p_{a}}^{p}\partial_{p_{b}}^{p}\right\} \mathscr{F}\left(x,p\right)\mathscr{G}\left(x,p\right)=\mathscr{F}\left(x,p\right)\mathscr{G}\left(x,p\right)\\
+\sum_{n=1}\frac{1}{n!}\left(\frac{\textrm{i}}{2}\right)^{n}\Theta^{a_{1}b_{1}}...\Theta^{a_{n}b_{n}}\partial_{a_{1}}^{x}...\partial_{a_{n}}^{x}\mathscr{F}\left(x,p\right)\partial_{b_{1}}^{x}...\partial_{b_{n}}^{x}\mathscr{G}\left(x,p\right)\\
+\sum_{n=1}\frac{1}{n!}\left(\frac{\textrm{i}}{2}\right)^{n}\eta^{a_{1}b_{1}}...\eta^{a_{n}b_{n}}\partial_{a_{1}}^{p}...\partial_{a_{n}}^{p}\mathscr{F}\left(x,p\right)\partial_{b_{1}}^{p}...\partial_{b_{n}}^{p}\mathscr{G}\left(x,p\right),
\end{array}\label{eq:4}
\end{equation}
with $\mathscr{F}\left(x,p\right)$ and $\mathscr{G}\left(x,p\right)$
assumed to be infinitely differentiable. If considering only the NC
space configuration, the $\star$product definition considerably reduces
to \citep{key-36,key-48}
\begin{equation}
\begin{array}{l}
\left(\mathscr{F}\star\mathscr{G}\right)\left(x\right)=\exp\left\{ \frac{\textrm{i}\Theta_{ab}}{2}\partial_{x_{a}}^{x}\partial_{x_{b}}^{x}\right\} \mathscr{F}\left(x\right)\mathscr{G}\left(x\right)=\mathscr{F}\left(x\right)\mathscr{G}\left(x\right)\\
+\sum_{n=1}\frac{1}{n!}\left(\frac{\textrm{i}}{2}\right)^{n}\Theta^{a_{1}b_{1}}...\Theta^{a_{n}b_{n}}\partial_{a_{1}}^{x}...\mathscr{F}\left(x\right)\partial_{b_{1}}^{x}...\partial_{b_{n}}^{x}\mathscr{G}\left(x\right).
\end{array}\label{eq:5}
\end{equation}

It should be emphasized that the spectral problem defined by the $\star$product
can be translated into an ordinary quantum mechanical problem via
the Bopp-shift. In the limit $\Theta=\eta=0$, the NC algebra reduces
to the standard Heisenberg algebra
\begin{equation}
\begin{array}{ccc}
\left[x_{i},x_{j}\right]=\left[p_{i},p_{j}\right]=0, & \textrm{and }\left[x_{i},p_{j}\right]=\textrm{i}\hbar\delta_{ij}, & \quad i,j=1,2\end{array}.\label{eq:6}
\end{equation}

\subsection{{\normalsize\label{II-B}}Gauge-Invariant Graphene in NCPS}

We first summarize the fundamental aspects of graphene physics in
an NCPS. In the commutative phase-space, the graphene system is governed
by the time-dependent Dirac equation
\begin{equation}
\mathscr{H}_{D}\psi=\textrm{i}\hbar\frac{\partial}{\partial t}\psi,\label{eq:7}
\end{equation}
where $\mathscr{H}_{D}$ consists of two massless Dirac Hamiltonians
describing electron dynamics near the two inequivalent Dirac points
(valleys) $K$ and $K'$. These points correspond to the corners of
the first Brillouin zone and are represented by the vectors \citep{key-2}
$\overrightarrow{K}=\frac{4\pi}{3\sqrt{3}a}\overrightarrow{e}_{x}$
and $\overrightarrow{K}^{'}=\frac{-4\pi}{3\sqrt{3}a}\overrightarrow{e}_{x}$,
where $a$ is the graphene lattice constant. These two Dirac points
determine the low-energy electronic excitations of graphene. At low
energies, graphene behaves as a zero-gap semiconductor with effectively
relativistic quasiparticles. Accordingly, the present work is restricted
to monolayer graphene. The wave function is written as
\begin{equation}
\psi=\left(\begin{array}{c}
\psi_{K}\textrm{e}^{-\frac{\textrm{i}}{\hbar}E_{K}t}\\
\psi_{K^{'}}\textrm{e}^{-\frac{\textrm{i}}{\hbar}E_{K'}t}
\end{array}\right),\label{eq:8}
\end{equation}
and $\psi_{K\left(K^{'}\right)}$ denotes the two-component spinor
in the vicinity of the $K\left(K^{'}\right)$ valley. The effective
Hamiltonian takes the block-diagonal form
\begin{equation}
\mathscr{H}_{K\left(K^{'}\right)}=\left(\begin{array}{cc}
H_{K} & 0\\
0 & H_{K^{'}}
\end{array}\right)=v_{F}\left(\begin{array}{cc}
\overrightarrow{\sigma}\cdot\overrightarrow{p} & 0\\
0 & \overrightarrow{\sigma}^{\ast}\cdot\overrightarrow{p}
\end{array}\right),\label{eq:9}
\end{equation}
with $\overrightarrow{\sigma}^{\ast}=\left(\sigma_{x},-\sigma_{y},\sigma_{z}\right)$.
Here $\overrightarrow{\sigma}$ denotes the $\textrm{2\texttimes2}$
Pauli matrices, which are given by
\begin{equation}
\sigma_{1}=\alpha_{1}=\left(\begin{array}{cc}
0 & 1\\
1 & 0
\end{array}\right),\;\sigma_{2}=\alpha_{2}=\left(\begin{array}{cc}
0 & -\text{i}\\
\text{i} & 0
\end{array}\right),\label{eq:10}
\end{equation}
and $\overrightarrow{p}=-\textrm{i}\hbar\left(\partial_{x},\partial_{y},0\right)$
is the momentum operator, and $v_{F}$ is the Fermi velocity. The
latter replaces the speed of light $c$ in graphene and is typically
$v_{F}\simeq c/300=10^{6}\textrm{ms}^{-1}$ \citep{key-16}. Eq. (\ref{eq:7})
leads to
\begin{equation}
\mathscr{H}_{K\left(K^{'}\right)}\psi_{K,K^{'}}=E_{K\left(K^{'}\right)}\psi_{K,K^{'}},\label{eq:11}
\end{equation}
with eigenvalues \citep{key-14} $E_{K\left(K^{'}\right)}=\pm\hbar v_{F}\left|\overrightarrow{k}\right|$.
In contrast, bilayer graphene exhibits a different low-energy structure
characterized by an approximately quadratic (parabolic) dispersion
relation and is therefore beyond the scope of the present analysis.
In momentum space, the eigenfunctions $\psi^{k}$ and $\psi^{k'}$
at the two Dirac points take the form \citep{key-4}
\begin{equation}
\psi_{\pm}^{k}(\overrightarrow{k})=\frac{1}{\sqrt{2}}\left(\begin{array}{c}
\textrm{e}^{-\frac{\textrm{i}\varphi_{k}}{2}}\\
\pm\textrm{e}^{\frac{\textrm{i}\varphi_{k}}{2}}
\end{array}\right),\qquad\psi_{\pm}^{k'}(\overrightarrow{k})=\frac{1}{\sqrt{2}}\left(\begin{array}{c}
\textrm{e}^{\frac{\textrm{i}\varphi_{k}}{2}}\\
\pm\textrm{e}^{-\frac{\textrm{i}\varphi_{k}}{2}}
\end{array}\right),\label{eq:12}
\end{equation}
with $\varphi_{k}=\arctan\left(k_{y}/k_{x}\right)$ is the polar angle
of vector $\overrightarrow{k}$, with $k_{x}=k_{F}\cos\varphi_{k}$,
$k_{y}=k_{F}\sin\varphi_{k}$ and $k_{F}$ is the Fermi momentum.
The signs $\pm$ correspond to the conduction and valence bands, respectively.

Now, under an external homogeneous MF in the z-direction $\overrightarrow{B}=B\overrightarrow{e}_{z}$,
the momentum operator is modified as $\overrightarrow{p}\rightarrow\overrightarrow{p}-e\overrightarrow{A}$.
The vector potential, $\overrightarrow{A}=\frac{B}{2}\left(-y,x,0\right)$
is induced by the MF, which is taken in the symmetric gauge. The Hamiltonian
in the commutative case reads as
\begin{equation}
\mathscr{H}_{K\left(K^{'}\right)}=v_{F}\left(\begin{array}{cc}
\sigma_{i}\left\{ p_{i}+\frac{eB}{2}\epsilon_{ij}x_{j}\right\}  & 0\\
0 & \sigma_{i}^{\ast}\left\{ p_{i}+\frac{eB}{2}\epsilon_{ij}x_{j}\right\} 
\end{array}\right)\:i,j=1,2.\label{eq:13}
\end{equation}

Through Eq. (\ref{eq:11}), and employing an appropriate annihilation
and creation operators, the energy eigenvalues \citep{key-57} is
obtained as $E_{K\left(K^{'}\right)}=\pm\frac{\hbar v_{F}}{l_{B}}\sqrt{2n}$
with $n=0,1,2,...$where we have used $l_{B}=\sqrt{\frac{\hbar}{eB}}$,
which is the magnetic length. 

Before introducing the NC framework, we recall that the standard NC
quantum-mechanical treatment of charged massless fermions, based solely
on the Bopp-shift or the $\star$product, generally breaks gauge-invariance.
A consistent formulation requires an NC field-theoretic approach combining
the SW map with the $\star$product, leading to a manifestly gauge-invariant
commutative equivalent \citep{key-56}.  The construction of the
gauge-invariant NCPS formulation proceeds in two stages. First, the
SW map together with the $\star$product is employed within an NC
field-theoretic framework to preserve gauge-invariance in NC space.
Subsequently, the Bopp-shift transformation is used to extend the
formalism to the full NCPS. The procedure may be summarized schematically
as follows:
\[
\left\{ \left(x_{i},p_{i}\right);\mathscr{H}\right\} \xrightarrow[\textrm{QFT}]{\textrm{SW map}\textrm{ \& }\star_{x}\textrm{product}}\left\{ \left(x_{i}^{nc},p_{i}\right);\mathscr{H}^{\Theta}\right\} \xrightarrow[p\rightarrow p^{nc}]{\;\textrm{Bopp-shift }\;}\left\{ \left(x_{i}^{nc},p_{i}^{nc}\right);\mathscr{H}^{nc}\right\} 
\]

Following the approach of Ref. \citep{key-60}, the gauge-invariant
graphene Hamiltonian in NC space is obtained by applying the SW map
and the $\star$product to Eq. (\ref{eq:13}). Then, using Eq. (\ref{eq:2}),
the gauge-invariant graphene Hamiltonian in NCPS is given by
\begin{equation}
\begin{array}{ccl}
\mathscr{H}_{K}^{nc} & = & v_{F}\left\{ \left(\alpha_{1}p_{x}+\alpha_{2}p_{y}\right)\left(1+\frac{\Theta}{2l_{B}^{2}}\right)+\left(\alpha_{1}y-\alpha_{2}x\right)\left[\frac{\eta}{2\hbar}+\frac{\hbar}{2l_{B}^{2}}\left(1+\frac{\Theta}{2l_{B}^{2}}\right)\right]\right\} +\mathcal{O}\left(\eta\Theta,\Theta^{2},\eta^{2}\right),\end{array}\label{eq:14}
\end{equation}

Here, the calculations were to the first-order in $\Theta$ and $\eta$
(also terms containing $\Theta\eta$ are neglected). The Hamiltonian
obtained in Eq. (\ref{eq:14}) describes massless Dirac quasiparticles
in graphene subjected to an external MF within NCPS. Physically, the
MF quantizes the electronic motion into relativistic Landau levels,
while the NC deformation introduces additional corrections associated
with the underlying noncommuting structure of coordinates and momenta.
The parameters $\Theta$ and $\eta$ characterize spatial and momentum
NCy, respectively, and together effectively encode short-distance
quantum-geometry corrections beyond the ordinary commutative description.
In the this framework, both deformations modify the effective coupling
between momentum and magnetic contributions in the graphene Hamiltonian.

It is also important to emphasize that gauge-invariance plays a central
role in the physical consistency of graphene under electromagnetic
interactions. In conventional NC quantum mechanics, the direct implementation
of the Bopp-shift or Moyal product generally breaks gauge symmetry,
leading to ambiguities in the coupling between charged particles and
the electromagnetic field. The combined SW map and $\star$product
formalism employed here avoids this difficulty and ensures that the
resulting effective Hamiltonian preserves the gauge structure of the
original graphene system.

Now, employing Eq. (\ref{eq:10}), Eq. (\ref{eq:14}) can be rewritten
in more compact form as follows:
\begin{equation}
\mathscr{H}_{K}^{nc}=v_{F}\left(\begin{array}{cc}
0 & \lambda\left\{ p_{x}-\textrm{i}p_{y}\right\} +\Omega\left\{ y+\textrm{i}x\right\} \\
\lambda\left\{ p_{x}+\textrm{i}p_{y}\right\} +\kappa\left\{ y-\textrm{i}x\right\}  & 0
\end{array}\right),\label{eq:15}
\end{equation}
with
\begin{equation}
\lambda=1+\frac{\Theta}{2l_{B}^{2}},\;\Omega=\frac{\hbar}{2l_{B}^{2}}\lambda+\frac{\eta}{2\hbar},\label{eq:16}
\end{equation}
where $\lambda$ represents the correction induced by coordinate NCy
to the kinetic sector of the Hamiltonian, while $\Omega$ combines
both magnetic and momentum-space NC contributions and acts as an effective
magnetic coupling parameter. Consequently, the NC deformation modifies
the effective cyclotron dynamics of the graphene quasiparticles.

The resulting Hamiltonian (\ref{eq:15}) preserves the Dirac-like
structure of graphene but with renormalized couplings generated by
the NCPS deformation. This deformation changes the spacing between
relativistic Landau levels and therefore modifies the density of accessible
quantum states. Introducing the following complex formalism:
\begin{equation}
z=x+\textrm{i}y,\;\overline{z}=x-\textrm{i}y,\label{eq:17}
\end{equation}
\begin{equation}
\begin{array}{cc}
p_{z}=-\textrm{i}\hbar\frac{d}{dz}=\frac{1}{2}\left(p_{x}-\textrm{i}p_{y}\right), & p_{\overline{z}}=-\textrm{i}\hbar\frac{d}{d\overline{z}}=\frac{1}{2}\left(p_{x}+\textrm{i}p_{y}\right),\end{array}\label{eq:18}
\end{equation}
 satisfying
\begin{equation}
\left[z,p_{z}\right]=\left[\bar{z},p_{\overline{z}}\right]=\textrm{i}\hbar,\;\left[z,p_{\overline{z}}\right]=\left[\bar{z},p_{z}\right]=0,\label{eq:19}
\end{equation}
Eq. (\ref{eq:15}) becomes
\begin{equation}
\mathscr{H}_{K}^{nc}=v_{F}\left(\begin{array}{cc}
0 & 2\lambda p_{z}+\Omega\textrm{i}\overline{z}\\
2\lambda p_{\overline{z}}-\Omega\textrm{i}z & 0
\end{array}\right).\label{eq:20}
\end{equation}

We now introduce the following creation and annihilation operators
\begin{equation}
\begin{cases}
a=\textrm{i}\left(\frac{\lambda}{\sqrt{\lambda\Omega\hbar}}p_{\overline{z}}-\frac{\textrm{i}}{2\lambda}\sqrt{\frac{\lambda\Omega}{\hbar}}z\right),\\
a^{\dagger}=-\textrm{i}\left(\frac{\lambda}{\sqrt{\lambda\Omega\hbar}}p_{z}+\frac{\textrm{i}}{2\lambda}\sqrt{\frac{\lambda\Omega}{\hbar}}\overline{z}\right),
\end{cases}\label{eq:21}
\end{equation}
satisfying the commutations relations $\left[a,a^{\dagger}\right]=1$,
$\left[a,a\right]=\left[a^{\dagger},a^{\dagger}\right]=0$. The state
$\left|\psi^{K}\right\rangle $ can be generated using the creation
operator as:
\begin{equation}
\left|\psi^{K}\right\rangle \equiv\left|n\right\rangle =\frac{\left(a^{\dagger}\right)^{n}}{n!}\left|0\right\rangle .\label{eq:22}
\end{equation}

Eq. (\ref{eq:20}) can be expressed in terms of creation and annihilation
operators (\ref{eq:21}) as:
\begin{equation}
\mathscr{H}_{K}^{nc}=\left(\begin{array}{cc}
0 & Qa^{\dagger}\\
Q^{*}a & 0
\end{array}\right),\label{eq:23}
\end{equation}
with $Q=2\textrm{i}v_{F}\sqrt{\lambda\Omega\hbar}$ determines the
effective strength of the coupling between neighboring Landau levels
in the NCPS. Let us return to the Dirac equation
\begin{equation}
\mathscr{H}_{K}^{nc}\left|\psi^{K}\right\rangle =E_{K}^{nc}\left|\psi^{K}\right\rangle ,\textrm{ with }\left|\psi^{K}\right\rangle =\left(\begin{array}{c}
\left|\phi_{\textrm{A}}\right\rangle \\
\left|\phi_{\textrm{B}}\right\rangle 
\end{array}\right),\label{eq:24}
\end{equation}
where $E_{K}^{nc}$ and $\left|\psi^{K}\right\rangle $ are the eigenvalues
and eigenkets of the NC Hamiltonian, respectively. Inserting Eq. (\ref{eq:23})
in Eq. (\ref{eq:24}), yields the following system of equations:
\begin{equation}
-E_{K}^{nc}\left|\phi_{\textrm{A}}\right\rangle +Qa^{\dagger}\left|\phi_{\textrm{B}}\right\rangle =0,\label{eq:25}
\end{equation}
\begin{equation}
Q^{*}a\left|\phi_{\textrm{A}}\right\rangle -E_{K}^{nc}\left|\phi_{\textrm{B}}\right\rangle =0.\label{eq:26}
\end{equation}

From Eq. (\ref{eq:25}), we have $\left|\phi_{\textrm{B}}\right\rangle =\frac{E_{K}^{nc}}{Qa^{\dagger}}\left|\phi_{\textrm{A}}\right\rangle ,$
subsequently, using Eq.(\ref{eq:26}) in the eigenbasis of the number
operator $a^{\dagger}a=N$, we obtain
\begin{equation}
\left\{ QQ^{*}N-\left(E_{K}^{nc}\right)^{2}\right\} \left|\phi_{\textrm{A}}\right\rangle =0,\,\mbox{with}\;N\left|\phi_{\textrm{A}}\right\rangle =n\left|\phi_{\textrm{A}}\right\rangle .\label{eq:27}
\end{equation}

Accordingly, the eigenenergy is
\begin{equation}
E_{K}^{nc}\left(n\right)=\pm\sqrt{4v_{F}^{2}\hbar\lambda\Omega n}.\label{eq:28}
\end{equation}

Using the explicit expression of $\lambda$ and $\Omega$, Eq. (\ref{eq:28})
can be rewritten as 
\begin{equation}
E_{K}^{nc}\left(n\right)=\pm\frac{\hbar v_{F}}{l_{B}}\sqrt{\left\{ 1+\bar{\Theta}\right\} \left\{ 1+\bar{\Theta}+\bar{\eta}\right\} 2n},\;\textrm{with }n=0,1,2,...\label{eq:29}
\end{equation}
with $\bar{\Theta}=\frac{\Theta}{2l_{B}^{2}}$ and $\bar{\eta}=\eta\frac{l_{B}^{2}}{\hbar^{2}}$
\citep{key-58}  being dimensionless constants related to the NC
parameter. The result (\ref{eq:29}) notably reveals that the landau
levels are altered by phase- and space-NCy. Positive values correspond
to electrons (conduction band), while negative ones stand for holes
(valence band). In the limit of $\Theta=\eta=0$, Eq. (\ref{eq:29})
reduces to the well-known commutative results \citep{key-57}, confirming
the physical consistency of the gauge-invariant NC formulation. By
considering Eqs. (\ref{eq:25}, \ref{eq:26}), and substitute Eq.
(\ref{eq:29}), we get the corresponding wave function for the Dirac
point $K$
\begin{equation}
\left|\psi_{k}\right\rangle ^{\textrm{T}}=\left(\begin{array}{cc}
\left|\phi_{A}\right\rangle  & \pm\textrm{i}\left|\phi_{B}\right\rangle \end{array}\right).\label{eq:30}
\end{equation}

The eigenvectors (\ref{eq:30}) can be written in polar coordinates
as follows \citep{key-81}:
\begin{equation}
\phi^{L}\left(n\right)=\mathcal{F}\left(\rho\right)\textrm{e}^{\textrm{i}m\varphi}\textrm{e}^{-\sigma\rho^{2}},\label{eq:31}
\end{equation}
where $L=\textrm{A},\textrm{B}$, and $\sigma>0$ is some constant.
$\rho=\sqrt{x^{2}+y^{2}}$, $\mathcal{F}\left(\rho\right)$ is some
polynomial of $\rho$, and $m$ is the quantum number associated to
the angular momentum $L_{z}$. 

Using the same method, the eigensystem of the Dirac point $K^{'}$
can be obtained. 

Now, we highlight the connection between the deformed graphene model
and the well-known Jaynes--Cummings (JC) and anti--Jaynes--Cummings
(AJC) models in quantum optics. The Hamiltonian (\ref{eq:23}) possesses
the same algebraic structure as the JC/AJC interaction models, a correspondence
between Dirac-like systems and quantum-optical models widely used
in graphene--quantum optics analogies \citep{key-82,key-83}. Using
Eq. (\ref{eq:10}), the Hamiltonian (\ref{eq:23}) can be exactly
rewritten as
\begin{equation}
\mathscr{H}_{K}=Q\left(\sigma^{+}a^{\dagger}+\sigma^{-}a\right),\label{eq:32-1}
\end{equation}
which corresponds to the AJC model, while the JC model is described
by the form \citep{key-84} $\mathscr{H}_{JC}=Q\left(\sigma^{+}a+\sigma^{-}a^{\dagger}\right)$.
Here,
\begin{equation}
\sigma^{\pm}=\frac{1}{2}\left(\sigma_{1}\pm i\sigma_{2}\right),\label{eq:33-1}
\end{equation}
denote the pseudospin raising and lowering operators. The JC model
\citep{key-85}, originally introduced to describe the interaction
between a two-level atom and a quantized electromagnetic field mode,
represents one of the fundamental exactly solvable models in quantum
optics. In contrast, the AJC model \citep{key-86} describes counter-rotating
interaction terms, where atomic excitation is accompanied by bosonic
excitation. Experimental realizations of these models, as well as
the connection between relativistic quantum mechanics and quantum
optics, have been extensively investigated in optical and microwave
cavities \citep{key-87}, trapped ions \citep{key-88}, and superconducting
circuits \citep{key-89}.  In this analogy, the graphene pseudospin
plays the role of two-level system, whereas the Landau-level ladder
operators behave analogously to bosonic photon operators. The NCPS
deformation modifies the effective coupling strength through the parameter
$Q$ without altering the underlying AJC algebraic structure. As a
result, the gauge-invariant graphene system in NCPS can be interpreted
as an effective deformed AJC quantum-optical model, where NC effects
introduce corrections to the coupling strength and energy spectrum
while preserving the fundamental interaction structure. 

\section{{\normalsize\label{III}}Thermal properties of gauge-invariant graphene
in NCPS}

\subsection{{\normalsize\label{III-A}}Thermodynamic Functions}

We explore the effect of phase-space NCy on the thermal properties
of the graphene system. In statistical physics, we use the partition
function $Z\left(\beta\right)$ to determine these properties. Given
the energy spectrum of our system $E_{n}$, the partition function
can be defined by a sum over all possible states $n$ of the system,
through
\begin{equation}
Z\left(\beta\right)=\sum_{n=0}^{\infty}\textrm{e}^{-\beta E_{n}},\textrm{ with }\beta=\frac{1}{k_{B}T},\label{eq:32}
\end{equation}
where $\beta$ is the Boltzmann factor, $T$ represents the equilibrium
temperature, and $K_{B}$ is the Boltzmann constant. Therefore, key
thermal properties such as free energy $F$, internal energy $U$,
entropy $S$, and specific heat capacity $C$ can be systematically
computed with the help of the following relations:
\begin{equation}
\begin{cases}
F=-\frac{\ln Z}{\beta},\;U=-\frac{\partial\ln Z}{\partial\beta},\\
S=\ln Z-\beta\frac{\partial\ln Z}{\partial\beta}=k_{B}\beta^{2}\frac{\partial F}{\partial\beta},\textrm{ and }C=\beta^{2}\frac{\partial^{2}\ln Z}{\partial\beta^{2}}=-k_{B}\beta^{2}\frac{\partial U}{\partial\beta}.
\end{cases}\label{eq:33}
\end{equation}

We revisit the spectrum (\ref{eq:29}), thus, it can be given as follows:
\begin{equation}
E_{K}^{nc}\left(n\right)=\pm\frac{\hbar v_{F}\sqrt{2}}{l_{B}}\sqrt{An},\label{eq:34}
\end{equation}
with
\begin{equation}
A=\left(1+\bar{\Theta}\right)\left(1+\bar{\Theta}+\bar{\eta}\right),\label{eq:35}
\end{equation}
which is the effective deformation factor encoding the combined contribution
of coordinate- and momentum-space NCy. In this part of our analysis,
we confine ourselves to the positive-energy stationary solutions.
The inclusion of negative-energy states leads to a non-convergent
(divergent) partition function within the canonical ensemble, thereby
preventing a consistent thermodynamic interpretation. Note that the
effective Dirac description of graphene admits an exact Foldy--Wouthuysen
transformation \citep{key-90}, which ensures a complete separation
between the positive- and negative-energy sectors without mixing.
As a result, these sectors evolve independently, and only the positive-energy
states contribute meaningfully to the thermodynamic properties of
the deformed graphene system. Accordingly, using the energy spectrum
given in Eq. (\ref{eq:34}), the single deformed graphene partition
function takes the form
\begin{equation}
Z=\sum_{n=0}^{\infty}\textrm{e}^{-\bar{\beta}\sqrt{An}},\label{eq:36}
\end{equation}
with \citep{key-58}:
\begin{equation}
\bar{\beta}=\frac{1}{\tau},\textrm{ and }\tau=\frac{T}{T_{0}}=\frac{l_{B}}{\hbar v_{F}\sqrt{2}}\frac{1}{\beta},\label{eq:37}
\end{equation}
where $\tau$ is the reduced temperature. Note that $T_{0}$ is the
characteristic temperature (its nature is very similar to the Debay
temperature) that divides the range of temperature to very high temperature,
$T\gg T_{0}$, and very low temperature, $T\ll T_{0}$ regions \citep{key-78}.
 Following the formula \citep{key-91,key-92}
\begin{equation}
\textrm{e}^{-x}=\frac{1}{2\pi\textrm{i}}\int_{c}\textrm{d}sx^{-s}\Gamma\left(s\right),\label{eq:38}
\end{equation}
Eq. (\ref{eq:36}) becomes
\begin{equation}
\sum_{n=0}^{\infty}\textrm{e}^{-\bar{\beta}\sqrt{An}}=\frac{1}{2\pi\textrm{i}}\int_{c}\textrm{d}s\left(\bar{\beta}\sqrt{A}\right)^{-s}\sum_{n}\left(n\right)^{-\frac{s}{2}}\Gamma\left(s\right)=\frac{1}{2\pi\textrm{i}}\int_{c}\textrm{d}s\left(\bar{\beta}\sqrt{A}\right)^{-s}\zeta\left(\frac{s}{2}\right)\Gamma\left(s\right),\label{eq:39}
\end{equation}
where $x=\bar{\beta}\sqrt{An}$, and $\Gamma\left(s\right)$ and $\zeta\left(\frac{s}{2}\right)$
are the Euler and Riemann zeta function, respectively. Using the residues
theorem, for the two poles $s=0$ and $s=2$, the partition function
is expressed in terms of the Riemann zeta function as follows
\begin{equation}
Z\left(\tau\right)=1+\frac{\tau^{2}}{A}+\zeta\left(0\right),\label{eq:40}
\end{equation}
with $\zeta_{H}\left(0\right)=-\frac{1}{2}$. Now, using the explicit
expression of $A$ in Eq. (\ref{eq:35}), the partition function in
NCPS is
\begin{equation}
Z\left(\tau,\bar{\Theta},\bar{\eta}\right)=\frac{1}{2}+\frac{\tau^{2}}{\left(1+\bar{\Theta}\right)\left(1+\bar{\Theta}+\bar{\eta}\right)}.\label{eq:41}
\end{equation}

According to Eq. (\ref{eq:41}), the partition function depends on
the NC parameters. Alternatively, the partition function can be evaluated
using the Euler--Maclaurin approximation, as shown in Annex \ref{A}.
Moreover, the deformed graphene system partition function can be easily
extended a system of $N$ distinguishable particle without internal
interactions through
\begin{equation}
\mathcal{Z}\left(\tau,\bar{\Theta},\bar{\eta}\right)=Z^{N}\left(\tau,\bar{\Theta},\bar{\eta}\right).\label{eq:42}
\end{equation}

Thermal quantities in Eq. (\ref{eq:33}) can be reformulated in terms
$\tau$ as follows:
\begin{equation}
\begin{cases}
\overline{F}=\frac{F}{\kappa}=-\tau\ln Z,\\
\overline{U}=\frac{U}{\kappa}=\tau^{2}\frac{\partial\ln Z}{\partial\tau},\\
\overline{S}=\frac{S}{k_{B}}=\ln Z+\tau\frac{\partial\ln Z}{\partial\tau},\\
\overline{C}=\frac{C}{k_{B}}=2\tau\frac{\partial\ln Z}{\partial\tau}+\tau^{2}\frac{\partial^{2}\ln Z}{\partial\tau^{2}},
\end{cases}\label{eq:43}
\end{equation}
with
\begin{equation}
\kappa=\frac{\hbar v_{F}\sqrt{2}}{l_{B}}\textrm{ and }\tau=\frac{1}{\kappa\beta},\label{eq:44}
\end{equation}
in which $\kappa$ defines the characteristic relativistic Landau
energy scale. The results demonstrate that the NCPS significantly
influences both the partition function and the thermal properties.
To further examine the behavior of these results, we present graphical
illustrations that highlight the impact of deformation effects on
the thermal properties.

\subsection{{\normalsize\label{III-B}Results} and Discussion}

Before presenting the plots, we first examine the asymptotic behavior
of the mean energy and the specific heat in the asymptotic limits.
In the high-temperature regime, $T\gg T_{0}$ (or equivalently $\tau\gg1$),
Eq. (\ref{eq:41}) can be approximated by
\begin{equation}
Z\simeq\frac{\tau^{2}}{A},\label{eq:45}
\end{equation}
from which the asymptotic behavior of the mean energy $U$ and the
specific heat $C$ follows:
\begin{equation}
\frac{U}{\kappa}\simeq2\tau,\text{ and }\frac{C}{K_{B}}\simeq2.\label{eq:46}
\end{equation}

This result may be understood by observing that the obtained limits
comply with the Dulong--Petit law for an ultra-relativistic ideal
gas, which shows that the NC corrections become subleading at sufficiently
high temperature. In contrast, in the low-temperature regime, $T\ll T_{0}$
(or $\tau\ll1$), the partition function reduces to
\begin{equation}
Z\simeq\frac{1}{2},\label{eq:47}
\end{equation}
consequently, both the mean energy and the specific heat vanish in
this limit, $U\rightarrow0$, $C\rightarrow0$, which is consistent
with the freezing of thermal excitations near the ground state.

To better understand the role of NCy, it is useful to analyze the
commutative limit of the model. In the commutative limit, $\bar{\Theta},\bar{\eta}\rightarrow0$,
one recovers $A\rightarrow1$, and the partition function reduces
to
\begin{equation}
Z_{\textrm{Comm}}=\frac{1}{2}+\tau^{2}.\label{eq:50-1}
\end{equation}
$\;$$\;$$\;$$\;$Accordingly, the reduced thermodynamic quantities
become
\begin{equation}
\begin{cases}
\overline{F}_{\textrm{Comm}}=-\tau\ln\left(\frac{1}{2}+\tau^{2}\right),\\
\overline{U}_{\textrm{Comm}}=\frac{2\tau^{3}}{\frac{1}{2}+\tau^{2}},\\
\overline{S}_{\textrm{Comm}}=\ln\left(\frac{1}{2}+\tau^{2}\right)+\frac{4\tau^{2}}{1+2\tau^{2}},\\
\overline{C}_{\textrm{Comm}}=\frac{3\tau^{2}+2\tau^{4}}{\left(\frac{1}{2}+\tau^{2}\right)^{2}},
\end{cases}\label{eq:51-1}
\end{equation}
which correspond to the standard thermodynamic behavior of graphene
Landau levels \citep{key-93}. 

\begin{figure}[H]
\centering{}%
\begin{tabular}{c}
\includegraphics[scale=0.7]{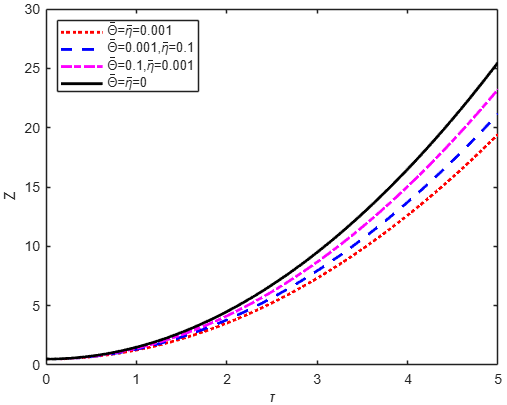}\tabularnewline
\end{tabular}\caption{\label{1}Partition function $Z$ of the gauge-invariant deformed
graphene as a function of $\tau$ for different values of $\bar{\Theta}$
and $\bar{\eta}$. }
\end{figure}

\begin{figure}[H]
\centering{}%
\begin{tabular}{cc}
\includegraphics[scale=0.63]{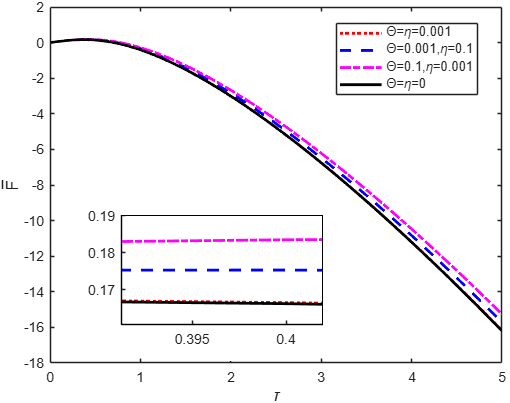} & \includegraphics[scale=0.63]{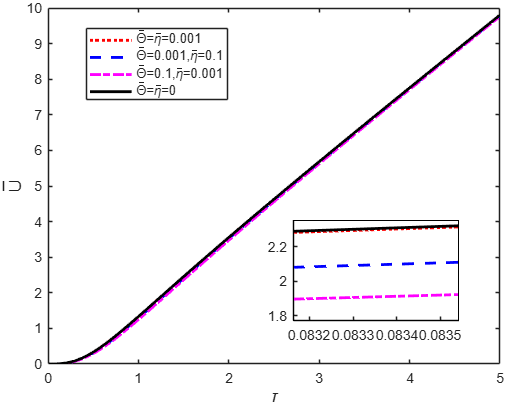}\tabularnewline
\includegraphics[scale=0.63]{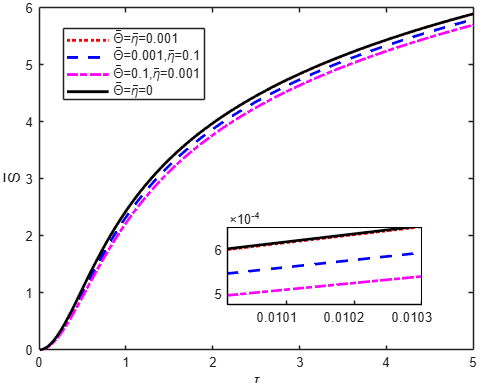} & \includegraphics[scale=0.63]{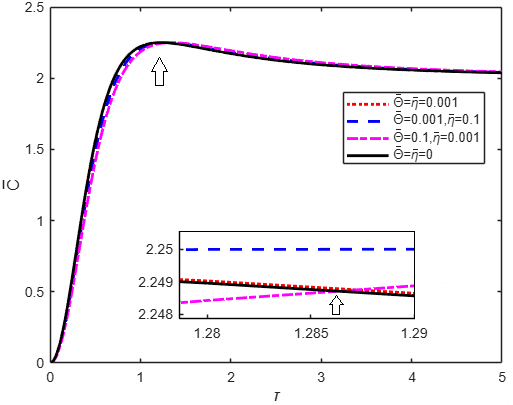}\tabularnewline
\end{tabular}\caption{\label{2}Reduced thermal quantities of the gauge-invariant deformed
graphene, namely $\overline{F}$, $\overline{U}$, $\overline{S}$,
and $\overline{C}$, as functions of $\tau$ for different values
of $\bar{\Theta}$ and $\bar{\eta}$. }
\end{figure}

Based on the obtained numerical results, we present plots illustrating
the thermodynamical behavior and the partition function of gauge-invariant
deformed graphene under variations of temperature $\left(\beta\left(\textrm{or }\tau\right)\right)$,
and the deformation parameters $\left(\Theta,\eta\right)$. Throughout
the numerical analysis we adopt $l_{B}=\hbar=K_{B}=1$. Several bounds
on the magnitudes of the NC parameters have been reported in the literature.
For instance, constraints of the order $\sqrt{\eta}\lesssim2.26\textrm{ \textmu eV}/c$
\citep{key-53} and $\sqrt{\Theta}\lesssim1\textrm{ TeV }$\citep{key-94}
have been obtained. In our graphical analysis, we therefore identify
the interval $0\prec\bar{\eta},\bar{\Theta}\prec0.1$. These estimates
are compatible with typical energy scales and experimental conditions
relevant to graphene systems. This range is physically meaningful
regime where NC effects may produce observable corrections. Within
this region the thermodynamic quantities show measurable deviations
from the commutative case. The partition function (Eq. (\ref{eq:41}))
is depicted in Fig. \ref{1}, while the reduced thermodynamical quantities
(Eq. (\ref{eq:43})), $\overline{F}$, $\overline{U}$, $\overline{S}$,
and $\overline{C}$ are plotted in Figs. \ref{2}. 

In Fig. \ref{1}, the partition function $Z$ of gauge-invariant deformed
graphene is plotted as a function of $\tau$ for different values
of $\bar{\Theta}$ and $\bar{\eta}$, and the results show that $Z$
increases monotonically with the parameter $\tau$, following a non-linear
growth pattern. The black solid line ($\bar{\Theta}=\bar{\eta}=0$)
represents the standard, undeformed state and serves as the upper
bound for the system's statistical weight. As the deformation parameters
$\bar{\Theta}$ and $\bar{\eta}$ increase, the curves shift downward,
indicating that these deformations suppress the partition function.
Specifically, the sensitivity of $Z$ to $\bar{\eta}$ appears more
pronounced than its sensitivity to $\bar{\Theta}$ in this regime,
with the combined presence of both parameters resulting in the lowest
values for $Z$. 

Fig. \ref{2} illustrates the behavior of reduced thermodynamic properties
for gauge-invariant deformed graphene as a function of the parameter
$\tau$ for different values of $\bar{\Theta}$ and $\bar{\eta}$.
In all cases, the deformation parameters $\bar{\Theta}$ and $\bar{\eta}$
cause noticeable shifts relative to the undeformed baseline (black
solid line), with $\overline{F}$ decreasing monotonically, with small
but noticeable deviations induced by the deformation parameters, particularly
in the low-temperature region as highlighted in the inset. Both $\overline{U}$
and $\overline{S}$ increasing as $\tau$ grows, with the entropy
shows the expected monotonic growth with temperature, with slight
parameter-dependent shifts. Notably, the heat capacity $\overline{C}$
exhibits a characteristic peak before stabilizing, a common feature
in finite systems or those with specific energy level densities. The
inset magnifications reveal that the deformation effects are subtle
but distinct, with the blue dashed line ($\bar{\eta}=0.1$) often
showing a more significant divergence than the magenta dashed line
($\bar{\Theta}=0.1$), suggesting the system's thermal stability and
state occupancy are more sensitive to $\bar{\eta}$ in this regime.

For completeness and to better visualize the dependence of the thermodynamic
functions on the deformation parameters, we also display plots over
the extended interval $\bar{\Theta},\bar{\eta}\in\left[0,10\right]$,
which is not intended to represent realistic physical values. However,
in this enlarged domain the NC corrections are appreciable only near
the origin ($\lesssim0.1$), while for larger values, the NC effects
become indistinguishable.

\begin{figure}[H]
\centering{}%
\begin{tabular}{c}
\includegraphics[scale=0.78]{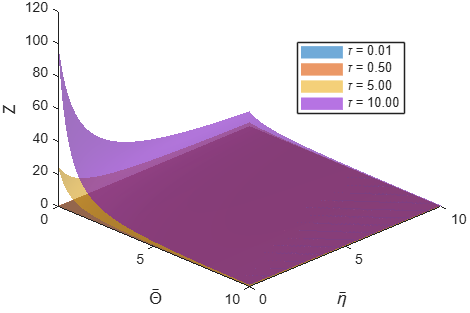}\tabularnewline
\end{tabular}\caption{\label{3}Partition function $Z$ of the gauge-invariant deformed
graphene vs. $\bar{\Theta}$ and $\bar{\eta}$ for different values
of $\tau$. }
\end{figure}

\begin{figure}[H]
\centering{}%
\begin{tabular}{cc}
\includegraphics[scale=0.7]{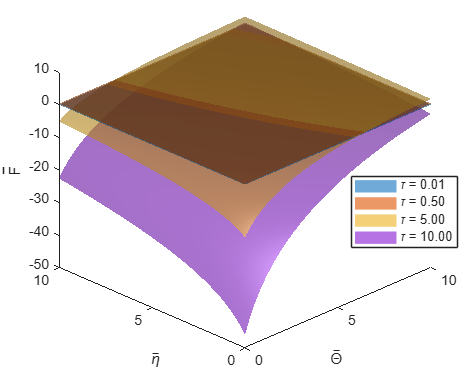} & \includegraphics[scale=0.7]{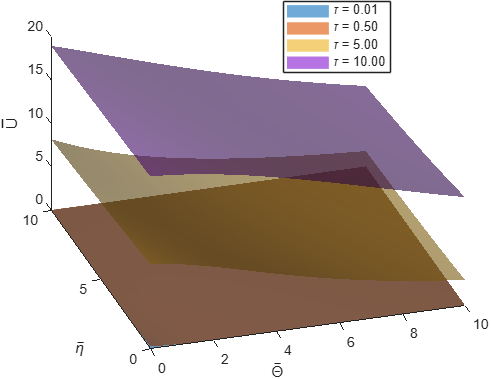}\tabularnewline
\includegraphics[scale=0.7]{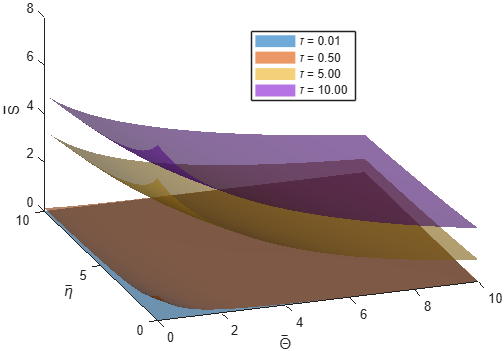} & \includegraphics[scale=0.7]{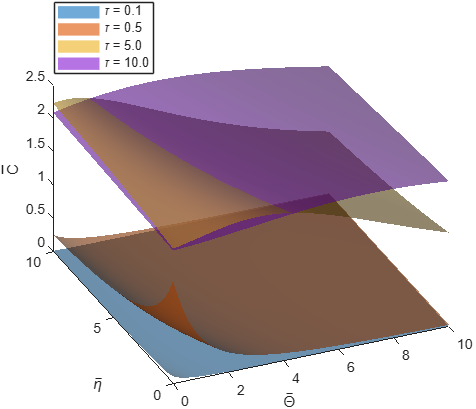}\tabularnewline
\end{tabular}\caption{\label{4}Reduced thermal quantities of the gauge-invariant deformed
graphene, namely $\overline{F}$, $\overline{U}$, $\overline{S}$,
and $\overline{C}$, vs. $\bar{\Theta}$ and $\bar{\eta}$ for different
values of $\tau$.}
\end{figure}

Fig. \ref{3} extend this analysis, showing the dependence of $Z$
on $\bar{\Theta}$ and $\bar{\eta}$ for different values of $\tau$.
We also plot $\overline{F}$, $\overline{U}$, $\overline{S}$, and
$\overline{C}$ as functions of $\bar{\Theta}$ and $\bar{\eta}$
for different values of $\tau$ in Fig. \ref{4}. 

The 3D surface plot in Fig. \ref{3} demonstrates the sensitivity
of the partition function to the NC parameters across various $\tau$
scales. It is evident that $Z$ reaches its maximum values when both
NC parameters approach zero, followed by a sharp, exponential-like
decay as either $\bar{\Theta}$ and $\bar{\eta}$ increases. This
suppression is most dramatic at higher values of $\tau$ (e.g., $\tau=10$,
represented by the purple surface), where the system's statistical
weight is significantly more sensitive to the deformation. Conversely,
for low $\tau$ values, the surfaces appear nearly flat and close
to the base, indicating that the impact of NC geometry on the partition
function is heavily dependent on the magnitude of the $\tau$ parameter.


The 3D surface plots in Fig. \ref{4} illustrate the dependence of
the reduced thermal quantities on the NC parameters across different
$\tau$ regimes. The Helmholtz free energy $\overline{F}$ shows a
significant downward curvature as both deformation parameters increase,
with the effect becoming more pronounced at higher $\tau$ values
(e.g., $\tau=10$). Conversely, the internal energy $\overline{U}$
and entropy $\overline{S}$ generally decrease as the NC parameters
increase, suggesting that the deformation reduces the system's ability
to store energy and its degree of disorder. The heat capacity $\overline{C}$
surfaces reveal a complex landscape where the impact of $\bar{\Theta}$
and $\bar{\eta}$ varies depending on whether the system is at a low
or high $\tau$ value, highlighting that the NC geometry significantly
alters the thermal stability and phase-like behavior of the gauge-invariant
deformed graphene. 

\begin{figure}[H]
\centering{}%
\begin{tabular}{cc}
\includegraphics[scale=0.6]{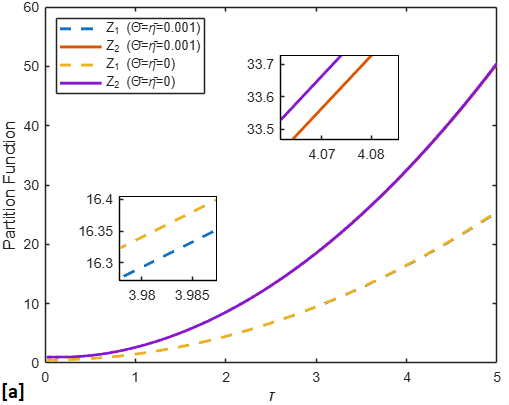} & \includegraphics[scale=0.6]{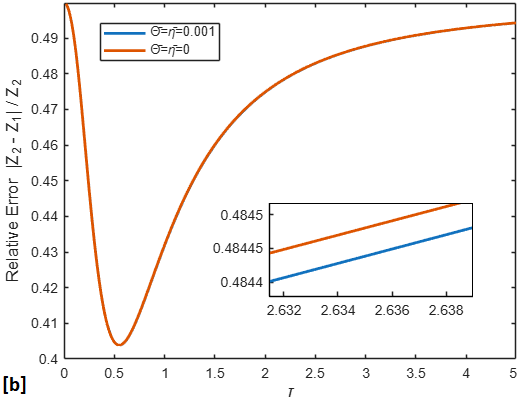}\tabularnewline
\end{tabular}\caption{\label{5}(a) Comparison of partition functions $Z_{1}$ (zeta) and
$Z_{2}$ (Euler--Maclaurin); (b) Log-Scale deviation between Euler--Maclaurin
and zeta results Vs. $\tau$, for commutative and NC cases.}
\end{figure}

It is worth now emphasizing that the two forms of the partition function
stem from different approximation schemes---the Riemann zeta function
and the Euler--Maclaurin expansion. Thus, a comparison of these partition
function formulas is shown in Fig. \ref{5}.

Fig. \ref{5} illustrates the divergence between the Riemann Zeta
and Euler--Maclaurin $Z$ formalisms across varying temperatures
$\tau$. In sub-fig. (\ref{5}.a), it is evident that while both expressions
coincide at the extreme low-temperature limit, they deviate significantly
as temperature increases; the Euler--Maclaurin result grows much
more rapidly, capturing a more complete set of thermal excitations
than the Riemann approximation. Sub-fig (\ref{5}.b) quantifies this
discrepancy through the relative error, revealing that the deviation
is not monotonic. It features a distinct local minimum near $\tau\approx0.5$
before rising and saturating at high temperatures, where the error
remains substantial (approaching $\approx50\%$). The insets further
show that the introduction of NC parameters ($\bar{\Theta}=\bar{\eta}=0.001$)
slightly shifts the partition function and reduces the relative error
compared to the commutative case, suggesting that the NCy subtly modulates
the convergence behavior of these two approximation schemes.

The Riemann zeta expression is compact and captures the dominant high-temperature
($\tau>1$) behavior, yielding a smooth quadratic growth with temperature
in which the NC parameters enter only through the effective deformation
factor $A$. In this limit, the discreteness of the Landau spectrum
and low-temperature corrections are negligible. By contrast, the Euler--Maclaurin
formulation approximates the level sum by an integral plus corrections,
producing an exponential Boltzmann factor multiplied by powers of
$\tau$ and $1/\tau$ terms. The exponential factor reflects the thermal
activation of Landau levels, while the $1/\tau$ contributions encode
finite-temperature quantum corrections, making this expression accurate
over a broader temperature range, particularly at low $\tau$. In
short, the Riemann zeta form provides the leading high-temperature
limit useful for analytical derivations, whereas the Euler--Maclaurin
expansion offers a more complete description suitable for tracking
the crossover from low to high temperatures.

Overall, one may conclude that the NC deformation modifies the effective
Landau scale and can be interpreted as an effective rescaling of the
Landau-level spacing, $E_{K}^{nc}\left(n\right)\propto\sqrt{An}$.
Hence, for $A>1$, the spacing between adjacent Landau levels increases,
making thermal excitation toward higher states more difficult. Physically,
this reduces the thermal population of excited states and therefore
suppresses the statistical contribution of higher Landau levels to
the partition function. As a consequence, the partition function and
entropy decrease compared with the commutative case, while the free
energy becomes more negative. Similarly, the specific heat exhibits
a suppression at intermediate temperatures because the enlarged Landau-level
spacing delays the thermal activation of excited states. The characteristic
peak structure of the specific heat remains present, reflecting the
discrete relativistic Landau spectrum. This demonstrates that the
NC parameters effectively modify the thermal accessibility of the
graphene quasiparticle states. The deviations from the commutative
thermodynamic behavior therefore originate directly from the deformation
of the relativistic Landau spectrum induced by NCPS effects in the
presence of the external MF. These modifications are not anomalous
but represent a consistent consequence of the NC-induced corrections
to the graphene energy structure. Despite the NC deformation, the
thermodynamic quantities preserve the expected physical behavior,
including positive specific heat, monotonic entropy behavior, and
vanishing thermal excitations in the zero-temperature limit. Therefore,
the low-temperature regime remains physically consistent.

It is instructive to place the present results within the broader
context of deformed graphene models and their thermodynamic properties.
Graphene in NCPS under an external MF was first investigated in Ref.
\citep{key-57}, where only momentum NCy was considered due to gauge-invariance
constraints associated with spatial NCy in standard NC quantum mechanics.
That study showed that NC effects do not modify the anomalous quantum
Hall effect and remain consistent with the fundamental physics of
graphene, while providing nontrivial corrections in the low-energy
regime. Subsequently, a variety of generalized deformed graphene frameworks
have been developed, including dynamical NC graphene \citep{key-61},
fractional and energy-dependent deformed graphene models \citep{key-60},
as well as formulations based on curved Snyder space \citep{key-70}
and Dunkl operators \citep{key-95}. Collectively, these approaches
highlight graphene as a versatile relativistic condensed-matter platform
for exploring different types of geometric and algebraic deformations.
In the context of thermodynamic properties, graphene has been previously
studied using the 2D Dirac oscillator formalism in Ref. \citep{key-93}.
In Refs. \citep{key-58,key-59}, based on the momentum-NC formulation
of Ref. \citep{key-57}, the thermal and magnetic properties of graphene
in NCPS were analyzed. However, those approaches neglected spatial
NCy, and therefore did not fully incorporate phase-space deformation
effects. In contrast, the present work provides a gauge-invariant
formulation that includes both spatial and momentum NC contributions
simultaneously. Moreover, the partition function used in Ref. \citep{key-58}
exhibits a divergence in the low-temperature limit ($\tau\rightarrow0$),
whereas the present formulation remains regular and yields consistent
asymptotic behavior in both low- and high-temperature regimes.

It is also useful to compare the present results with other deformed
frameworks scenarios. In Ref. \citep{key-69}, graphene thermodynamics
was studied within Doubly Special Relativity, particularly in the
Amelino--Camelia and Magueijo--Smolin models, where modified dispersion
relations lead to significant changes in the partition function and
specific heat, indicating that thermodynamic observables can distinguish
between different relativistic deformations. Similarly, energy-dependent
NC graphene \citep{key-60} exhibits asymmetric modifications of electron
and hole Landau levels, leading to electron--hole symmetry breaking
and spectral bifurcations not present in the current NCPS framework.
Fractional graphene model \citep{key-60} introduce fractional Landau
spectra with modified Dirac dynamics, while the position-dependent
translation operator (PDTO) formalism \citep{key-71} yields effective
mass and Fermi-velocity renormalization, together with modified Landau-level
spacing, while largely preserving particle--hole symmetry. In curved
Snyder space \citep{key-70}, it has been shown that the impact on
graphene thermodynamics becomes significant mainly in the high-temperature
regime.

Overall, these studies demonstrate that different deformation frameworks
lead to qualitatively distinct modifications at both the spectral
and thermodynamic levels. In the present gauge-invariant NC model,
the deformation primarily acts through a rescaling of the effective
Landau-level spacing, which directly modifies the partition function
and reduces the thermal accessibility of excited states. 

Overall, these comparisons indicate that different deformation frameworks
affect graphene in qualitatively distinct ways, both at the spectral
and thermodynamic levels. In the present gauge-invariant NCPS model,
the deformation primarily modifies the effective Landau-level spacing,
which in turn directly impacts the partition function and the thermal
accessibility of excited states. This leads to systematic deviations
in the thermodynamic quantities compared with the commutative case,
while preserving physically consistent low- and high-temperature limits.
In contrast to other deformation schemes, where effects such as electron--hole
asymmetry or anomalous scaling may dominate, the NCPS framework produces
a controlled and symmetric modification of the thermodynamic response
governed by the combined phase-space structure. These results further
support the view that graphene constitutes a natural laboratory for
exploring NC geometry and more general deformed quantum frameworks
through their measurable thermodynamic signatures.

\section{{\normalsize\label{IV}Conclusion}}

In this work, we investigated the thermal behavior of a gauge-invariant
graphene system within NCPS framework under a constant external MF.
Since standard NC formulations generally lead to gauge-invariance
issues, a consistent gauge-invariant construction of the graphene
Hamiltonian in NCPS was first established. Using the ladder-operator
formalism, the corresponding deformed Landau levels and eigenstates
were derived. On this basis, the canonical partition function was
computed using the Riemann zeta-function method, allowing the determination
the thermodynamic quantities, namely the free energy, internal energy,
entropy, and specific heat capacity. The resulting thermodynamic behavior,
presented in Figs. \ref{2}--\ref{4}, shows that phase-space NCy
induces systematic deviations from the commutative graphene case.
The curves of the mean energy and specific heat exhibit excellent
agreement with the asymptotic limits given in Eq. (\ref{eq:46}).
In addition, a mapping of gauge-invariant deformed model to the AJC
model was established, providing a bridge to quantum-optical interpretation. 

We suggest that 2D Dirac materials such as graphene provide a promising
platform for probing NC geometries and for exploring possible experimental
signatures. Possible extensions of this work include the study of
anomalous quantum Hall effect, Zitterbewegung dynamics, and the investigation
of possible emergence of a quantum phase transition in this framework.

\section*{Funding}

This research received no external funding. 

\section*{Declarations Conflict of interest }

The author declares no conflict of interest.

\appendix

\section{\label{A}Partition function using Euler-Maclaurin formula}

The single deformed graphene partition function (\ref{eq:36}) can
be presented as
\begin{equation}
Z=1+\sum_{n=0}^{\infty}\textrm{e}^{-\bar{\beta}\sqrt{An+A}},\label{eq:A1}
\end{equation}

We evaluate the partition function of deformed graphene employing
the Euler-Maclaurin formula. Prior to this, we examine the convergence
of the partition function (\ref{eq:A1}) using the integral test.
We consider the integral $\mathscr{B}$ of the function $f\left(x\right)=\exp\left(-\frac{1}{\tau}\sqrt{Ax+A}\right)$,
which is a positive, monotonically decreasing function. The corresponding
integral is given by
\begin{equation}
\mathscr{B}\left(\tau\right)=\int_{0}^{+\infty}\textrm{e}^{-\frac{1}{\tau}\sqrt{Ax+A}}dx=\frac{2\tau^{2}}{A}\left(1+\frac{1}{\tau}\sqrt{A}\right)\textrm{e}^{-\frac{\sqrt{A}}{\tau}},\label{eq:A2}
\end{equation}
which is finite. thereafter, according to the integral test, the numerical
partition function $Z$ is convergent. We now employ the Euler-Maclaurin
formula, defined as follows \citep{key-96,key-97}:
\begin{equation}
\sum_{n=0}^{\infty}f\left(n\right)=\frac{1}{2}f\left(0\right)+\int_{0}^{+\infty}f\left(x\right)dx-\sum_{p=1}^{\infty}\frac{B_{2p}}{\left(2p\right)!}f^{(2p-1)}\left(0\right),\label{eq:A3}
\end{equation}
where $f^{(2p-1)}\left(0\right)$ denotes the derivative of order
$(2p-1)$ at $x=0$, and the coefficients $B_{2p}$ are Bernoulli
numbers, expressed as \citep{key-96}
\begin{equation}
B_{n}=\frac{2\left(2n\right)!}{\left(2\pi\right)^{2n}}\sum_{p=1}^{\infty}p^{-2n},\label{eq:A4}
\end{equation}
and key specific values are $B_{2}=1/6$, $B_{4}=-1/30$, and $B_{6}=1/42$.
We calculate the sum in Eq. (\ref{eq:A6}) using Eq. (\ref{eq:A4}),
so, up to $p=2$ we obtain 
\begin{equation}
\sum_{p=1}^{\infty}\frac{B_{2p}}{\left(2p\right)!}f^{(2p-1)}\left(0\right)=\textrm{e}^{-\frac{\sqrt{A}}{\tau}}\frac{\sqrt{A}}{\tau}\left(\frac{79}{1920}-\frac{1}{1920}\frac{\sqrt{A}}{\tau}-\frac{1}{5760}\frac{A}{\tau^{2}}\right).\label{eq:A5}
\end{equation}

Hence, substituting Eq. (\ref{eq:A5}) in Eq. (\ref{eq:A3}), the
partition function is explicitly obtained as follows:
\begin{equation}
Z\left(\tau\right)=1+\frac{1}{2}\textrm{e}^{-\frac{\sqrt{A}}{\tau}}+\frac{2\tau^{2}}{A}\textrm{e}^{-\frac{\sqrt{A}}{\tau}}\left(1+\frac{\sqrt{A}}{\tau}\right)-\textrm{e}^{-\frac{\sqrt{A}}{\tau}}\frac{\sqrt{A}}{\tau}\left(\frac{79}{1920}-\frac{\sqrt{A}}{1920\tau}-\frac{A}{5760\tau^{2}}\right).\label{eq:A6}
\end{equation}

Then, using the explicit expression of $A$, the partition function
in NCPS is
\begin{equation}
\begin{array}{c}
Z\left(\tau,\bar{\Theta},\bar{\eta}\right)=1+\textrm{e}^{-\frac{\sqrt{\left(1+\bar{\Theta}\right)\left(1+\bar{\Theta}+\bar{\eta}\right)}}{\tau}}\left\{ \frac{1}{2}+\frac{2}{\left(1+\bar{\Theta}\right)\left(1+\bar{\Theta}+\bar{\eta}\right)}\tau^{2}+\frac{2}{\sqrt{\left(1+\bar{\Theta}\right)\left(1+\bar{\Theta}+\bar{\eta}\right)}}\tau\right.\\
\left.-\frac{79\sqrt{\left(1+\bar{\Theta}\right)\left(1+\bar{\Theta}+\bar{\eta}\right)}}{1920}\frac{1}{\tau}+\frac{\left(1+\bar{\Theta}\right)\left(1+\bar{\Theta}+\bar{\eta}\right)}{1920}\frac{1}{\tau^{2}}+\frac{\left\{ \left(1+\bar{\Theta}\right)\left(1+\bar{\Theta}+\bar{\eta}\right)\right\} ^{3/2}}{5760}\frac{1}{\tau^{3}}\right\} .
\end{array}\label{eq:A7}
\end{equation}

The Euler--Maclaurin summation formula provides a useful bridge between
discrete sums and continuum integrals. In the present problem, the
thermodynamic quantities involve infinite sums over Landau levels,
and the Euler--Maclaurin approach allows one to approximate these
sums analytically while preserving the dominant thermal contributions.
Physically, this approximation becomes particularly relevant in the
high-temperature regime, where many excited Landau levels contribute
to the partition function and the spectrum behaves quasi-continuously.
The consistency between the Euler--Maclaurin approximation and the
zeta-function evaluation further supports the validity of the obtained
thermodynamic expressions.


\begin{thebibliography}{10}
\bibitem{key-1}Choi, W., et al. Synthesis of Graphene and Its Applications:
A Review. Crit. Rev. Solid. State. Mater. Sci. \textbf{35}(1), 52
(2010). \url{https://doi.org/10.1080/10408430903505036}

\bibitem{key-2}Fefferman, C., Weinstein, M. Honeycomb lattice potentials
and Dirac points. J. Amer. Math. Soc. \textbf{25}, 1169 (2012). \url{https://doi.org/10.1090/S0894-0347-2012-00745-0}

\bibitem{key-3}Novoselov, K.S., et al. \href{https://doi.org/10.1126/science.1102896}{Electric Field in Atomically Thin Carbon Films}.
Science, \textbf{306}, 666 (2004). 

\bibitem{key-4}Castro Neto, A.H., et al. \href{https://doi.org/10.1103/RevModPhys.81.109}{The electronic properties of graphene}.
Rev. Mod. Phys. \textbf{81}, 109 (2009). 

\bibitem{key-5}Peres, N.M.R. The transport properties of graphene.
J. Phys.: Condens. Matter. \textbf{21}, 323201 (2009). 

\bibitem{key-6}Peres, N.M.R. Colloquium: The transport properties
of graphene: An introduction. Rev. Mod. Phys. \textbf{82} (2010) 2673.
\url{https://doi.org/10.1103/RevModPhys.82.2673}

\bibitem{key-7}Min, K. Aluru, N.R. Mechanical properties of graphene
under shear deformation. Appl. Phys. Lett. \textbf{98}(1), 013113
(2011). \url{https://doi.org/10.1063/1.3534787}

\bibitem{key-8}Falkovsky, L.A. \href{https://doi.org/10.1088/1742-6596/129/1/012004}{Optical properties of graphene}.
J. Phys.: Conf. Ser. \textbf{129}(1), 012004 (2008). 

\bibitem{key-9}Pop, E., et al. Thermal properties of graphene: Fundamentals
and applications. MRS Bull. \textbf{37}(12), 1273 (2012). \url{https://doi.org/10.1557/mrs.2012.203}

\bibitem{key-10}Geim, A.K. \href{https://doi.org/10.1126/science.1158877}{Graphene: status and prospects}.
science, \textbf{324}(5934), 1530 (2009). 

\bibitem{key-11}Huang, X., et al. \href{https://doi.org/10.1039/C1CS15078B}{Graphene-based composites}.
Chem. Soc. Rev. \textbf{41}(2), 666 (2011). 

\selectlanguage{french}%
\bibitem{key-12}\foreignlanguage{english}{Weiss, N.O., et al. \href{https://doi.org/10.1002/adma.201201482}{Graphene: an emerging electronic material}.
Adv. Mater. \textbf{24}(43), 5782 (2012). }

\bibitem{key-13}\foreignlanguage{english}{Jiang, Z., et al. Infrared
Spectroscopy of Landau Levels of Graphene. Phys. Rev. Lett.\textbf{
98}, 197403, (2007). \url{https://doi.org/10.1103/PhysRevLett.98.197403}}

\bibitem{key-14}\foreignlanguage{english}{Geim, A., Novoselov, K.
\href{https://doi.org/10.1038/nmat1849}{The rise of graphene}. Nat.
Mater. \textbf{6}, 183 (2007). }

\bibitem{key-15}\foreignlanguage{english}{Bonaccorso, F., et al.
\href{https://doi.org/10.1038/nphoton.2010.186}{Graphene photonics and optoelectronics}.
Nat. Photon. \textbf{4}, 611 (2010). }

\bibitem{key-16}\foreignlanguage{english}{Goerbig, M.O. Electronic
properties of graphene in a strong magnetic field. Rev. Mod. Phys.
\textbf{83}, 1193 (2011). \url{https://doi.org/10.1103/RevModPhys.83.1193}}

\bibitem{key-17}\foreignlanguage{english}{Rusanov, A.I. \href{https://doi.org/10.1016/j.surfrep.2014.08.003}{Thermodynamics of graphene}.
Surf. Sci. Rep. \textbf{69}(4), 296 (2014). }

\bibitem{key-18}\foreignlanguage{english}{Katsnelson, M., et al.
Chiral tunnelling and the Klein paradox in graphene. Nat. Phys. \textbf{2},
620 (2006). \url{https://doi.org/10.1038/nphys384}}

\bibitem{key-19}\foreignlanguage{english}{Katsnelson, M.I., Novoselov,
K. S. Graphene: New bridge between condensed matter physics and QED.
Solid. State. Commun. \textbf{143}, 3 (2007). \url{https://doi.org/10.1016/j.ssc.2007.02.043}}

\selectlanguage{english}%
\bibitem{key-20} Katsnelson, M.I. \href{https://doi.org/10.1016/S1369-7021(06)71788-6}{Graphene: carbon in two dimensions}.
Mater. Today. \textbf{10}(1), 20 (2007). 

\selectlanguage{french}%
\bibitem{key-21}\foreignlanguage{english}{Bagarello, F., Hatano,
N. PT-symmetric graphene under a magnetic field. Proc. R. Soc. A.
\textbf{472}, 20160365 (2016).\url{http://dx.doi.org/10.1098/rspa.2016.0365}}

\bibitem{key-22}\foreignlanguage{english}{Kim, P. (2017). \href{https://doi.org/10.1007/978-3-319-32536-1_1}{Graphene and Relativistic Quantum Physics}.
In: Duplantier, B., Rivasseau, V., Fuchs, JN. (eds) Dirac Matter .
Progress in Mathematical Physics, vol 71. Birkhäuser, Cham. }

\selectlanguage{english}%
\bibitem{key-23}Cooper, D.R., D\textquoteright Anjou, B., et al.
Experimental review of graphene. Int. Sch. Res. Notices. \textbf{2012}(1),
501686 (2012). \url{https://doi.org/10.5402/2012/501686}

\selectlanguage{french}%
\bibitem{key-24}\foreignlanguage{english}{Gusynin, V.P., Sharapov,
S.G. Unconventional integer quantum Hall effect in graphene. Phys.
Rev. Lett. \textbf{95}(14), 146801 (2005). \url{https://doi.org/10.1103/PhysRevLett.95.146801}}

\selectlanguage{english}%
\bibitem{key-25}Snyder , H. S. \href{https://doi.org/10.1103/PhysRev.71.38}{Quantized Space-Time}.
Phys. Rev. \textbf{71}, 38 (1947). 

\bibitem{key-26}Snyder , H. S. \href{https://doi.org/10.1103/PhysRev.72.68}{The Electromagnetic Field in Quantized Space-Time}.
Phys. Rev. \textbf{72}, 68 (1947). 

\selectlanguage{french}%
\bibitem{key-27}\foreignlanguage{english}{Connes, A. Non-commutative
differential geometry. Publications Mathématiques de L\textquoteright Institut
des Hautes Scientifiques. \textbf{62}, 41 (1985). \url{https://doi.org/10.1007/BF02698807}}

\bibitem{key-28}\foreignlanguage{english}{Woronowicz, S. L. Twisted
SU(2) Group. An Example of a Non-Commutative Differential Calculus.
Publ. Res. Inst. Math. Sci. \textbf{23}(1), 117 (1987). \url{https://doi.org/10.2977/prims/1195176848}}

\bibitem{key-29}\foreignlanguage{english}{Woronowicz, S.L. \href{https://doi.org/10.1007/BF01219077}{Compact matrix pseudogroups}.
Commun. Math. Phys. \textbf{111}, 613 (1987). }

\bibitem{key-30}\foreignlanguage{english}{Haouam, I. On the noncommutative
geometry in quantum mechanics. J. Phys. Stud. \textbf{24}(2), 2002
(2020). \url{https://doi.org/10. 30970/jps.24.2002 }}

\bibitem{key-31}\foreignlanguage{english}{Carroll, S.M., et al. Noncommutative
field theory and Lorentz violation. Phys. Rev. Lett. \textbf{87}(14),
141601 (2001). \url{https://doi.org/10.1103/PhysRevLett.87.141601}}

\bibitem{key-32}Seiberg, N., Witten, E. String theory and noncommutative
geometry. J. High. Energy. Phys. \textbf{1999}(9), 032 (1999). \url{https://doi.org/10.1088/1126-6708/1999/09/032}

\selectlanguage{english}%
\bibitem{key-33}\foreignlanguage{french}{Connes, Ac et al. Noncommutative
geometry and matrix theory. J. High. Energy. Phys. \textbf{1998}(02),
003 (1998). \url{https://doi.org/10.1088/1126-6708/1998/02/003}}

\selectlanguage{french}%
\bibitem{key-34}\foreignlanguage{english}{Gingrich, D.M. Noncommutative
geometry inspired black holes in higher dimensions at the LHC. J.
High. Energy. Phys. \textbf{2010}, 22 (2010). \url{https://doi.org/10.1007/jhep05(2010)022}}

\bibitem{key-35}\foreignlanguage{english}{Gracia-Bondia, J.M. \href{https://doi.org/10.1007/978-3-642-11897-5_1}{Notes on Quantum Gravity and Noncommutative Geometry: New Paths Towards Quantum Gravity}
(Springer, Berlin, Heidelberg, 2010). }

\bibitem{key-36}\foreignlanguage{english}{Gangopadhyay, S., et al.
On the Landau system in noncommutative phase-space. Phys. Lett. A,
379, 2956--2961 (2015). \url{https://doi.org/10.1016/j.physleta.2015.08.039} }

\bibitem{key-37}\foreignlanguage{english}{Martinetti, P. \href{https://doi.org/10.1088/1742-6596/634/1/012001}{Beyond the standard model with noncommutative geometry, strolling towards quantum gravity}.
J. Phys.: Conf. Ser. \textbf{634} 012001 (2015). }

\selectlanguage{english}%
\bibitem{key-38}Bellissard, J., et al. The noncommutative geometry
of the quantum Hall effect. J. Math. Phys. \textbf{35}, 53731 (1994).
\url{https://doi.org/10.1063/1.530758}

\selectlanguage{french}%
\bibitem{key-39}\foreignlanguage{english}{Santos, W.O., Souza, A.M.C.
\href{https://doi.org/10.1007/s10773-012-1280-x}{The Anomalous Zeeman Effect for the Hydrogen Atom in Noncommutative Space}.
Int. J. Theor. Phys. \textbf{51}, 3882 (2012).  }

\bibitem{key-40}\foreignlanguage{english}{Falomir, H., et al. Testing
spatial noncommutativity via the Aharonov-Bohm effect. Phys. Rev.
D. 66, 045018 (2002). \url{https://doi.org/10.1103/PhysRevD.66.045018} }

\bibitem{key-41}\foreignlanguage{english}{Bastos, C., Bertolami,
O. The Berry Phase in the Noncommutative Gravitational Quantum Well.
Phys. Lett. A, \textbf{372}, 5556 (2008). \url{https://doi.org/10.1016/j.physleta.2008.06.073} }

\bibitem{key-42}\foreignlanguage{english}{Chaichian, M., et al. Hydrogen
atom spectrum and the Lamb shift in noncommutative QED. Phys. Rev.
Lett. \textbf{86} (2001) 2716. \url{https://doi.org/10.1103/PhysRevLett.86.2716} }

\bibitem{key-43}\foreignlanguage{english}{Haouam, I. Solutions of
Noncommutative Two-Dimensional Position--Dependent Mass Dirac Equation
in the Presence of Rashba Spin-Orbit Interaction by Using the Nikiforov--Uvarov
Method. Int. J. Theor. Phys. \textbf{62}, 111 (2023). \url{https://doi.org/10.1007/s10773-023-05361-5} }

\bibitem{key-44}\foreignlanguage{english}{Haouam, I. The Non-Relativistic
Limit of the DKP Equation in Non-Commutative Phase-Space. Symmetry,
\textbf{11}, 223 (2019). \url{https://doi.org/10.3390/sym11020223} }

\bibitem{key-45}\foreignlanguage{english}{Novikov, O.O. \ensuremath{\mathscr{P}}\ensuremath{\mathscr{T}}-symmetric
quantum field theory on the noncommutative spacetime. Mod. Phys. Lett.
A, \textbf{35}(05), 2050012 (2020). \url{https://doi.org/10.1142/S0217732320500121} }

\bibitem{key-46}\foreignlanguage{english}{Gáliková, V., et al. Quantum
Mechanics in Noncommutative Space. Acta. Physica. Slovaca, \textbf{65}(3),
153 (2015). \url{http://www.physics.sk/aps/pub.php?y=2015&pub=aps-15-03} }

\bibitem{key-47}\foreignlanguage{english}{Moslehian, M. S., Kian,
M. Non-commutative-divergence functional. Mathematische Nachrichten,
\textbf{286}(14-15), 1514 (2013). \url{https://doi.org/10.1002/mana.201200194} }

\bibitem{key-48}\foreignlanguage{english}{Haouam, I. Classical limit
and Ehrenfest\textquoteright s theorem versus non-relativistic limit
of noncommutative Dirac equation in the presence of minimal uncertainty
in momentum. Int. J. Theor. Phys, \textbf{62}, 189 (2023). \url{https://doi.org/10.1007/s10773-023-05444-3} }

\bibitem{key-49}\foreignlanguage{english}{Buric, M., et al. \href{https://doi.org/10.3842/SIGMA.2007.125}{WKB approximation in noncommutative gravity}.
SIGMA. \textbf{3}, 125 (2007). }

\bibitem{key-50}\foreignlanguage{english}{Najazadeh, M., Saadat,
M. Thermodynamics of Classical Systems on Noncommutative Phase Space.
Chin. J. Phys. \textbf{51}, 94 (2013). \url{https://doi.org/10.6122/CJP.51.94} }

\bibitem{key-51}\foreignlanguage{english}{Gouba, L. A comparative
review of four formulations of noncommutative quantum mechanics. Int.
J. Mod. Phys. A, \textbf{31}, 1630025 (2016). \url{https://doi.org/10.1142/S0217751X16300258} }

\bibitem{key-52}\foreignlanguage{english}{Li, K., et al. Representation
of noncommutative phase space. Mod. Phys. Lett. A. \textbf{20}(28),
2165 (2005). \url{https://doi.org/10.1142/S0217732305017421} }

\bibitem{key-53}\foreignlanguage{english}{Bertolami, O., Queiroz,
R. Phase-space noncommutativity and the Dirac equation. Phys. Lett.
A, \textbf{375}(46), 4116 (2011). \url{https://doi.org/10.1016/j.physleta.2011.09.053}}

\bibitem{key-54}\foreignlanguage{english}{Haouam, I. Ehrenfest's
theorem for the Dirac equation in noncommutative Phase-Space. Math.
Comput. Sci. \textbf{4}(4), 53-63 (2024). \url{https://doi.org/10.30511/mcs.2023.2013583.1143} }

\selectlanguage{english}%
\bibitem{key-55}\foreignlanguage{french}{Halder, A. \href{https://doi.org/10.1140/epjp/s13360-025-06842-8}{Noncommutative Landau problem in graphene: a gauge-invariant analysis with the Seiberg\textendash Witten map}.
Eur. Phys. J. Plus, \textbf{140}, 925 (2025).} 

\bibitem{key-56}Halder, A., et al. Gauge-invariant description of
a Dirac electron moving in a noncommutative gauge field background.
EPL, \textbf{149}, 60001 (2025). \url{https://doi.org/10.1209/0295-5075/adbe9e}

\bibitem{key-57}Bastos, C., et al. \href{https://doi.org/10.1142/S0217751X13500644}{Noncommutative graphene}.
Int. J. Mod. Phys. A, \textbf{28}(16), 1350064 (2013). 

\selectlanguage{french}%
\bibitem{key-58}\foreignlanguage{english}{Santos, V., et al. Thermodynamical
properties of graphene in noncommutative phase--space. Ann. Phys.
\textbf{349}, 402 (2014). \url{https://doi.org/10.1016/j.aop.2014.07.005}}

\bibitem{key-59}\foreignlanguage{english}{Khordad, R., Rastegar Sedehi,
H.R. \href{https://doi.org/10.1140/epjp/i2019-12558-5}{Magnetic susceptibility of graphene in non-commutative phase-space: Extensive and non-extensive entropy}.
Eur. Phys. J. Plus, \textbf{134}, 133 (2019). }

\bibitem{key-60}\foreignlanguage{english}{Haouam, I. Graphene in
Energy-Dependent Noncommutative Phase-Space. Next. Res.\textbf{7},
101483 (2026). \url{https://doi.org/10.1016/j.nexres.2026.101483}}

\selectlanguage{english}%
\bibitem{key-61}Haouam, I., Alavi, S. A. Dynamical noncommutative
graphene. Int. J. Mod. Phys. A, \textbf{37}(10), 2250054 (2022). \url{https://doi.org/10.1142/S0217751X22500543}

\bibitem{key-62}Dayi, Ö. F., Jellal, A. A noncommutative space approach
to confined Dirac fermions in graphene. J. Math. Phys. \textbf{51},
063522 (2010). \url{https://doi.org/10.1063/1.3442719}

\selectlanguage{french}%
\bibitem{key-63}\foreignlanguage{english}{Monemzadeh, M., et al.
Embedding of a 2D Graphene System in Non-Commutative Space. J. Nanostruct,
\textbf{3}(3), 315 (2013). \url{https://doi.org/10.7508/jns.2013.03.007}}

\bibitem{key-64}\foreignlanguage{english}{Falomir, H., et al. Graphene
and non-Abelian quantization. J. Phys. A: Math. Theor. \textbf{45}
135308 (2012). \url{https://doi.org/10.1088/1751-8113/45/13/135308}}

\bibitem{key-65}\foreignlanguage{english}{Taie, M., et al. BFT Embedding
and Gauge Symmetries of Graphene System in Non-Commutative Space.
Int. J. Theor. Phys. \textbf{54}, 2334 (2015). \url{https://doi.org/10.1007/s10773-014-2455-4}}

\bibitem{key-66}\foreignlanguage{english}{Sourrouille, L. Landau
levels for graphene layers in noncommutative plane. Int. J. Mod. Phys.
A, \textbf{36}(14), 2150087 (2021).\url{https://doi.org/10.1142/S0217751X21500871}}

\bibitem{key-67}\foreignlanguage{english}{ Hoc, N. Q., et al. Thermodynamic
properties of graphene in the presence of perpendicular magnetic and
in-plane electric fields with asymmetrical Gaussian confinement potentials.
Phys. Lett. A,\textbf{ 578}, 131473 (2026). \url{https://doi.org/10.1016/j.physleta.2026.131473}}

\bibitem{key-68}\foreignlanguage{english}{Gbetoho, J.M., et al. Thermodynamic
Properties of Graphene in an External Magnetic Field Within the Framework
of the Generalized Uncertainty Principle. J. Low. Temp. Phys. \textbf{222},
89 (2026). \url{https://doi.org/10.1007/s10909-026-03413-8}}

\bibitem{key-69}\foreignlanguage{english}{Boumali, A., Jafari, N.
\href{https://doi.org/10.1007/s10909-026-03376-w}{Thermal Properties of Graphene Via Dirac Oscillator Model Under Doubly Special Relativity Framework}.
J. Low. Temp. Phys. \textbf{222}, 46 (2026). }

\bibitem{key-70}\foreignlanguage{english}{Hamil, B., et al. \href{https://doi.org/10.1515/zna-2020-0159}{Graphene in curved Snyder space}.
Z. Naturforsch. A, \textbf{75}(10), 809-817 (2020). }

\bibitem{key-71}\foreignlanguage{english}{Aguiar, V., et al. \href{https://doi.org/10.1103/PhysRevB.102.235404}{Dirac fermions in graphene using the position-dependent translation operator formalism}.
Phys. Rev. B, \textbf{102}(23), 235404 (2020). }

\bibitem{key-72}\foreignlanguage{english}{Moures, N., et al. Schwinger
pair creation for monolayer graphene in a constant electromagnetic
field, and in noncommutative phase space coordinates. Int. J. Mod.
Phys. A, \textbf{38}(02), 2350008 (2023). \url{https://doi.org/10.1142/S0217751X23500082}}

\bibitem{key-73}\foreignlanguage{english}{Chen, K., et al. Application
of constrained forms of the Tsallis entropy in thermodynamic properties
of graphene based on modified Heisenberg model. Int. J. Geom. Methods
Mod. Phys, \textbf{23}(08), 2550230 (2026). \url{https://doi.org/10.1142/S0219887825502305}}

\bibitem{key-74}\foreignlanguage{english}{Rowe, P., et al. Development
of a machine learning potential for graphene. Phys. Rev. B, \textbf{97}(5),
054303 (2018). \url{https://doi.org/10.1103/PhysRevB.97.054303}}

\bibitem{key-75}\foreignlanguage{english}{Huang, M., et al. Recent
advances of graphene and related materials in artificial intelligence.
Adv. Intell. Syst, \textbf{4}(10), 2200077 (2022). \url{https://doi.org/10.1002/aisy.202200077} }

\bibitem{key-76}\foreignlanguage{english}{Singh, A., Li, Y. \href{https://doi.org/10.1016/j.commatsci.2023.112272}{Reliable machine learning potentials based on artificial neural network for graphene}.
Comput. Mater. Sci, \textbf{227}, 112272 (2023).}

\bibitem{key-77}\foreignlanguage{english}{Sabattini, L., et al. Towards
AI-driven autonomous growth of 2D materials based on a graphene case
study. Commun. Phys, \textbf{8}, 180 (2025). \url{https://doi.org/10.1038/s42005-025-02086-1}}

\selectlanguage{english}%
\bibitem{key-78}Haouam, I. Thermal behavior of the Klein Gordon oscillator
in a dynamical noncommutative space. Sci. Rep. \textbf{15}, 28771
(2025). \url{https://doi.org/10.1038/s41598-025-10118-7}

\bibitem{key-79}Iorio, A., et al. Turning graphene into a lab for
noncommutativity. Phys. Lett. B, \textbf{852}, 138630 (2024). \url{https://doi.org/10.1016/j.physletb.2024.138630}

\bibitem{key-80}Dey, S., Fring, A. Noncommutative quantum mechanics
in a time-dependent background. Phys. Rev. D, \textbf{90}(8), 084005
(2014). \url{https://doi.org/10.1103/PhysRevD.90.084005}

\bibitem{key-81}Tannoudji, C. C., Diu, B., Laloë, F. (1973). Mécanique
quantique. Collection Ensignement.

\selectlanguage{french}%
\bibitem{key-82}\foreignlanguage{english}{Karnieli, A., Fan, S. Jaynes-Cummings
interaction between low-energy free electrons and cavity photons.
Science advances, \textbf{9}(22), eadh2425 (2023). \url{https://doi.org/10.1126/sciadv.adh2425}}

\bibitem{key-83}\foreignlanguage{english}{Bocanegra-Garay, I. A.,
et al. \href{https://doi.org/10.1103/PhysRevResearch.6.043218}{Exploring supersymmetry: interchangeability between Jaynes-Cummings and anti-Jaynes-Cummings models}.
Phys. Rev. Res, \textbf{6}(4), 043218 (2024). }

\bibitem{key-84}\foreignlanguage{english}{Bermudez, A., et al. Chirality
quantum phase transition in the Dirac oscillator. Phys. Rev. A, \textbf{77}(6),
063815 (2008). \url{https://doi.org/10.1103/physreva.77.063815 }}

\bibitem{key-85}\foreignlanguage{english}{Jaynes, E., and Cummings,
F. \href{https://doi.org/10.1109/PROC.1963.1664}{Comparison of quantum and semiclassical radiation theories with application to the beam maser}.
Proc. IEEE 51, 89 (1963). }

\bibitem{key-86}\foreignlanguage{english}{Solano, E., et al. Strong-driving-assisted
multipartite entanglement in cavity QED. Phys. Rev. Lett. 90, 027903
(2003). \url{https://doi.org/10.1103/PhysRevLett.90.027903}}

\bibitem{key-87}\foreignlanguage{english}{M. Brune, F. et al. Quantum
Rabi oscillation: A direct test of field quantization in a cavity,
Phys. Rev. Lett. \textbf{76}, 1800 (1996). \url{https://doi.org/10.1103/PhysRevLett.76.1800}}

\bibitem{key-88}\foreignlanguage{english}{Bermudez, A., et al. \href{https://doi.org/10.1103/PhysRevA.76.041801}{Exact mapping of the 2+1 Dirac oscillator onto the Jaynes-Cummings model: Ion-trap experimental proposal}.
Phys. Rev. A 76, 041801 (2007). }

\bibitem{key-89}\foreignlanguage{english}{A. Blais, A. L., et al.
\href{https://doi.org/10.1103/RevModPhys.93.025005}{Circuit quantum electrodynamics}.
Rev. Mod. Phys. \textbf{93}, 025005 (2021). }

\selectlanguage{english}%
\bibitem{key-90}Silenko, A.J. \href{https://doi.org/10.1134/S1547477123050680}{Foldy\textendash Wouthuysen Transformation and Structured States of a Graphene Electron in External Fields and Free (2+1)-Space}.
Phys. Part. Nuclei Lett. \textbf{20}, 1131 (2023). 

\bibitem{key-91}Dariescu, M. A., Dariescu, C. \href{https://doi.org/10.1088/0953-8984/19/25/256203}{Persistent currents and critical magnetic field in planar dynamics of charged bosons}.
J. Phys. : Condens. Matter, \textbf{19}, 256203 (2007). 

\bibitem{key-92}Dariescu, M. A., Dariescu, C. \href{https://doi.org/10.1016/j.chaos.2006.03.021}{Finite temperature analysis of quantum Hall-type behavior of charged bosons}.
Chaos. Solitons and Fractals, \textbf{33}, 776--781 (2007). 

\selectlanguage{french}%
\bibitem{key-93}\foreignlanguage{english}{Boumali, A. \href{https://doi.org/10.1088/0031-8949/90/4/045702}{Thermodynamic properties of the graphene in a magnetic field via the two-dimensional Dirac oscillator}.
Phys. Scr.\textbf{ 90} (4). 90 045702 (2015). }

\bibitem{key-94}\foreignlanguage{english}{Bertolami, O., et al. Noncommutative
gravitational quantum well. Phys. Rev. D \textbf{72}, 025010 (2005).
\url{https://doi.org/10.1103/PhysRevD.72.025010}}

\bibitem{key-95}Hamil, B., Lütfüo\u{g}lu, B.C. Dunkl graphene in
constant magnetic field. Eur. Phys. J. Plus, \textbf{137}, 1241 (2022).
\url{https://doi.org/10.1140/epjp/s13360-022-03463-3}

\selectlanguage{english}%
\bibitem{key-96}G. Arfken, Mathematical Methods for Physicists, 3rd
ed. (Academic Press, Orlando, FL, 1985), pp. 327--338.

\bibitem{key-97}Pacheco, M. H., et al. One-dimensional Dirac oscillator
in a thermal bath. Phys. Lett. A. \textbf{311}, 93 (2003). \url{https://doi.org/10.1016/S0375-9601(03)00467-5}

\end{thebibliography}
\end{document}